\documentclass[preprint]{aastex}
\usepackage{emulateapj5}
\shorttitle{H$_2$O Masers of Cepheus A HW2}
\shortauthors{Gallimore et al.}

\received{2002 November 5}
\accepted{2002 November 20}

\begin{document}

\newcommand{\water}{H$_2$O}
\newcommand{\kms}{km~s$^{-1}$}
\newcommand{\mone}{$^{-1}$}

\title{Expansion of the R4 \water\ Maser Arc Near Cepheus A HW2}
\author{J. F. Gallimore}
\affil{Dept. of Physics, Bucknell University, Lewisburg, PA 17837}
\email{jgallimo@bucknell.edu}
\author{R. J. Cool}
\affil{Dept. of Physics \& Astronomy, P.O. Box 3905, University of
  Wyoming, Laramie, WY 82071}
\author{M. D. Thornley}
\affil{Dept. of Physics, Bucknell University, Lewisburg, PA 17837}
\and
\author{J. McMullin}
\affil{National Radio Astronomy Observatory, Socorro, NM 87801}

\begin{abstract}

We present new (April 2000) MERLIN observations of the \water\ masers
located near the protostar Cepheus A HW2. The MERLIN observations
detect many of the structures found in earlier (1996) VLBA
observations of Torrelles and collaborators, and the changed positions
of these structures are compatible with the VLBA proper motions and
astrometric uncertainties.  The radius of curvature of the R4
structure of maser arcs appears to have grown by a factor of two, and
the displacement of the arcs between 1996 and 2000 are compatible with
expansion about a common center. In addition, the MERLIN observations
detect red-shifted masers not previously found; taken with the newly
discovered masers, the R4 structure now resembles patchy emission from
an elliptical ring. We demonstrate that a simple bow-shock model
cannot simultaneously account for the shape and the velocity gradient
of the R4 structure. A model involving a slow, hydromagnetic shock
propagating into a rotating, circumstellar disk better describes the
maser spot kinematics and luminosities. In this model, the central
mass is  $3M_{\odot}  $, and we demonstrate that the mass of the disk is
negligible in comparison. The expansion velocity of the post-shock gas,  $\sim
5\,\mbox{km s}^{ - 1} $, is slow compared to the shock velocity,  $v_S
\sim 13\,\mbox{km s}^{ - 1} $, suggesting that the post-shock gas is
magnetically supported with a characteristic field strength of  $\sim
30 $ mG.  We speculate that the expanding maser rings R4 and R5 may be
generated by periodic, instability-driven winds from young stars
that periodically send spherical shocks into the surrounding circumstellar
material.

\end{abstract}


\section{Introduction}

Interstellar \water\ masers trace warm, dense molecular gas associated
with young stellar objects (YSOs) and star-forming regions (e.g.,
Genzel {\&} Downes 1977; more recent examples include Claussen et
al. 1998; Furuya et al. 2000; Rodr\'iguez et al. 2002).  In contrast
with thermal tracers, \water\ masers are bright, compact radio
sources, and can be detected on very long baselines, corresponding to
sub-milliarcsecond (mas) resolution. Using aperture synthesis
techniques, one can map both the geometry of the maser spots and trace
their very fine proper motions. Adding radial velocity information
derived from the Doppler shift of the maser line results in a detailed
picture of the maser geometry and kinematics as a measure of the
dynamics very close to the YSO (e.g., Genzel et al.  1981a; Genzel et
al. 1981b; Schneps et al.  1981; Gwinn, Moran, {\&} Reid 1992).

Most of the known interstellar masers trace gas located near the base
of molecular outflows on scales of 10s -- 100s of AU (for reviews, see
Reid {\&} Moran 1988; Elitzur 1992) The association of masers with
outflows has been unambiguous. The maser spots (i.e., individual,
unresolved maser sources) align with molecular outflows mapped in
millimeter-wave emission lines, or are associated with Herbig-Haro
objects. The kinematics of the such masers also follow the sense of
the motion of the outflow on larger scales. The proper motions are
usually small compared to the molecular outflow velocities, suggesting
that \water\ masers trace the expanding shock front between the
outflow and the ambient ISM (Claussen 2001).

There are, however, a handful of studied interstellar \water\ maser
sources that appear to be associated with circumstellar disks or rings
rather than molecular outflows (Matveenko 1987; Cesaroni 1990; Fiebig
et al. 1996; Torrelles et al. 1996; Berulis, Lekht, {\&}
Mendoza-Torres 1998; Torrelles et al. 1998a; Torrelles et al. 1998b;
Hunter et al. 1999; Shepherd {\&} Kurtz 1999; Matveenko {\&} Diamond
2000; Patel et al. 2000; Lekht {\&} Sorochenko 2001).  These disks are
thought to be the remnants of the molecular cloud cores out of which
the stars formed (e.g., Shu, Adams, {\&} Lizano 1987).  The nature of
these disks ultimately bears on the origin of planets, which are
generally thought to form out of condensations in the circumstellar
disk (e.g., as reviewed in McCaughrean 1997).

Recently, Torrelles and collaborators 
(Torrelles, et al.  1996; 
(Torrelles, et al. 1998a)
reported evidence for \water\ masers in a
protoplanetary disk around the continuum radio source Ceph A HW2
(Hughes {\&} Wouterloot 1984a). 
The radio continuum source resolves
into a jet of thermal free-free emission oriented along PA 44\degr\ 
(Rodriguez et al. 1994). 
Torrelles et al. used the VLA to map the
maser emission and found a compact distribution of spots, roughly 300
AU (0\farcs4) in extent (assuming a distance of 725 pc: 
Johnson 1957). 
The maser spots coarsely align at right angles to the jet axis,
i.e., are better aligned with the predicted position angle of a
protoplanetary disk than with the thermal jet. In addition, there is a
radial velocity gradient across the distribution of maser spots that
is compatible with disk rotation.

Their subsequent 1996 VLBA observations resolved the apparent disk
into a more complex, filamentary arrangement of maser spots. The maser
spot filaments (called R1 -- R5) show seemingly independent proper
motions and may well be associated with unseen protostars in the
neighborhood of HW2 instead of a large-scale disk associated solely
with HW2 
(Torrelles et al. 2001a; 
Torrelles et al. 2001b). 
In this work, we present new MERLIN observations of the \water\ masers
located near HW2. Our main result is the discovery of an expanding
ring of maser spots associated with region R4 (using the naming
convention of Torrelles et al.). There is a velocity gradient along
the major axis, which is most simply explained by rotation.  We
consider the possibility that the R4 masers arise from a gaseous disk
surrounding a forming star.

This paper is organized as follows. \S2 describes both the new MERLIN
and archival VLA observations of the HW2 \water\ masers, and \S3 presents
the results of these observations. \S4 describes two shock models for the HW2 R4-A maser
arcs and argues in favor of a model involving a shock wave propagating
into a rotating, gaseous disk. We also present some constraints on the
properties of the pre-shock gas based on the shock models. \S5
considers the proper motions of the other arcuate maser structures, R1
-- R5 with the goal of placing a limit on the proper motion of the
rotation center of the R4 masers. \S6 summarizes the results of the
observations and shock modeling and also briefly discusses the
possible origin of the R4 masers.

\begin{figure*}
\plotone{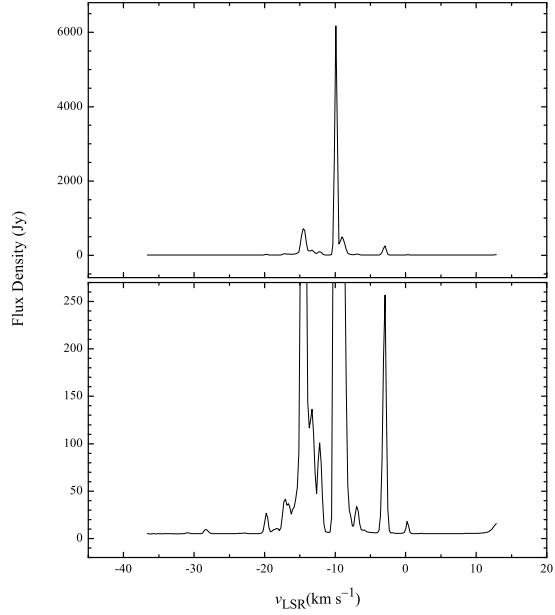}
\figcaption{Plots of the MERLIN total power spectrum of the \water\ masers
in the region of Ceph A HW2 and HW3. This spectrum was derived from
the calibrated visibility $(u, v)$ data rather than the channel
images. The top panel is the spectrum scaled to show the brightest
emission (which is associated with HW3), and the bottom panel is the
same spectrum scaled to emphasize the fainter masers.
\label{fig1}}
\end{figure*}

\section{Observations and Data Reduction}

\subsection{MERLIN Maser Data}

On 9 April 2000, we used the 5-element Multi-Element Radio-Linked
Interferometer Network (MERLIN\footnote{ MERLIN is operated by the
University of Manchester on behalf of the Particle Physics \&
Astronomy Research Council (PPARC)}) telescope, based at the Jodrell
Bank Observatory, to observe the Ceph A HW2 region (pointing center:
$\alpha $[1950] = 22$^{\mbox{h}} $ 54$^{\mbox{m}} $
19.0592$^{\mbox{s}} $, $\delta $[1950] = 61\degr 45\arcmin
46.51\arcsec). The central frequency was tuned to the 6$_{16} $ --
5$_{23} $ \water\ maser transition (rest frequency: 22235.079 MHz),
offset to the systemic velocity of the Ceph A region, $v_{LSR} = -12
$~\kms.  The observations were made in spectral line mode with 256
channels and a channel spacing of 0.21 km s$^{ - 1} $. The total
on-source integration time was 11.5 hrs.

Calibration involved initial editing and bandpass calibration using
the ``dprocs'' routines developed at Jodrell Bank. We then used AIPS
to compute an improved bandpass calibration, as well as atmospheric
and instrumental phase and amplitude corrections against the position
of the calibrator point source J2302+640, whose coordinates are known
to a precision of about 12 mas 
(Patnaik et al. 1992).  
The integrated
maser spectrum derived from the calibrated visibility data is plotted
in Figure~\ref{fig1}.

After calibration, the data were transformed into images of the sky
using the AIPS task IMAGR, which performs both the requisite Fourier
Transform and a ``CLEAN'' deconvolution 
(H\"{o}gbom 1974; Clark 1980). 
To achieve the optimum angular resolution, the visibility data
grid was given uniform weight during the Fourier Transform. The
resulting angular resolution (``clean beam'') was 8 mas circular.  
The RMS noise in signal-free channels is 25
mJy beam$^{ - 1}$, comparable to the expected thermal limit of  $\sim  $
20 mJy beam$^{ - 1}$. However, channels containing bright maser
emission were always dynamic range limited at a ratio of typically
400:1 (signal: noise).

\begin{figure*}
\plotone{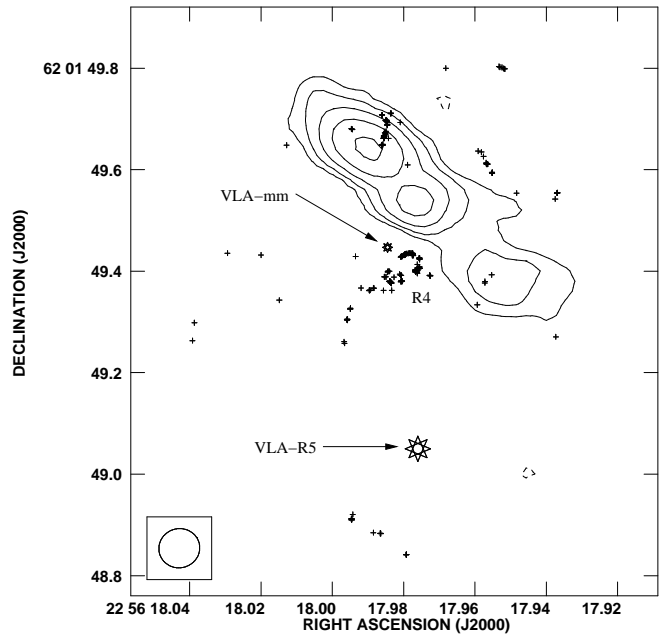}
\figcaption{Comparison of the MERLIN maser positions (small crosses) with
the 22 GHz VLA radio continuum image (contours). The contour levels
are 0.7, 1.2, 2.1, 3.5, and 6.0 mJy beam$^{-1} $. The VLA beam size,
plotted in the lower left corner, is 0.08\arcsec. The location of two
newly discovered, faint continuum sources, VLA-mm and VLA-R5 (Curiel
et al. 2002), are plotted as eight-point stars.
\label{fig2}}
\end{figure*}

\begin{figure*}
\plotone{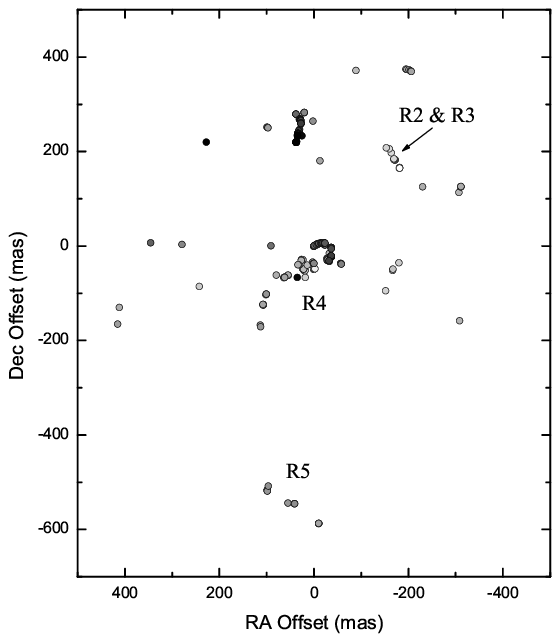}
\plotone{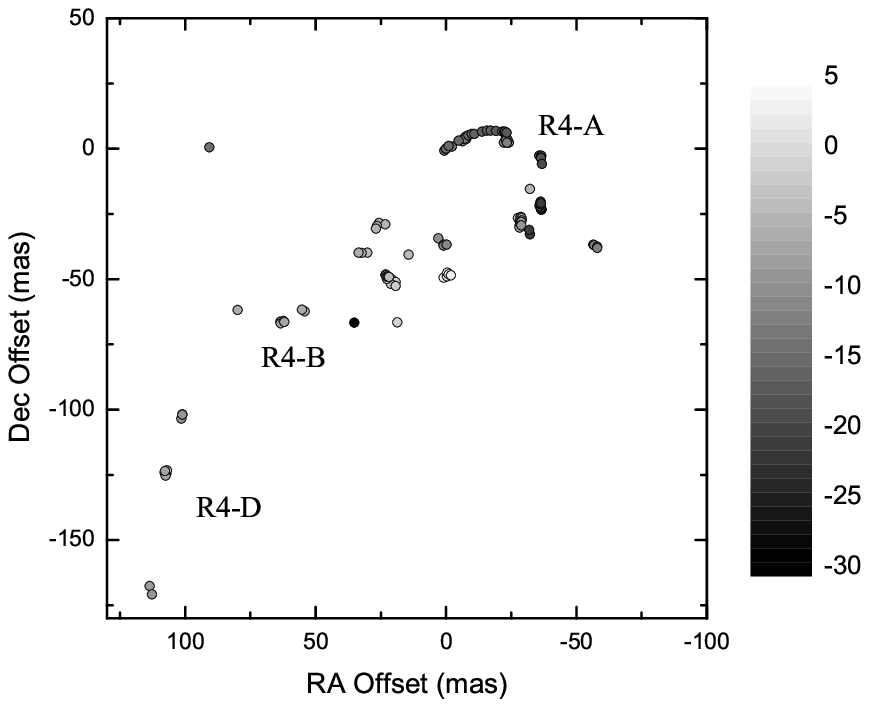}
\figcaption{Plot of the HW2 maser positions and velocities based on the
new MERLIN data. The axes are sky offset positions relative to the
brightest maser within the MERLIN data. Each spot is shaded according
to the radial velocity scale given in the bottom panel. Left panel:
large-scale field showing all of the MERLIN detections over the HW2
region. Right panel: close-up field showing the details of the R4
region. 
\label{fig3}}
\end{figure*}

Each channel image was then examined for maser signals of
signal-to-noise ratio (SNR) $> 5 $, from which we compiled a database
of maser positions and velocities. The positions of the MERLIN maser
spots are plotted in Figures~\ref{fig2} and \ref{fig3}. For
convenience, and unless otherwise specified, offset coordinates are
referenced to the position of the brightest spot associated with the
R4 maser arc. The reference position is $\alpha $(J2000) $ =
22^{\mathrm{h}} $ $56^{\mathrm{m}} $ 17\fs97792, Dec (J2000) =
$+62\degr $ 01\arcmin 49\farcs4421.


\subsection{VLA Continuum Data}

We also obtained archival VLA continuum observations of Ceph A HW2 to
compare with the new MERLIN \water\ maser data. The data were
originally obtained and published by 
Torrelles et al.(1998a). 
We re-reduced the data following standard techniques within the AIPS
software package. The flux calibration and astrometry of the resulting
image compare very well with those presented by Torrelles et al.

We converted the radio continuum image to J2000 coordinates using the
AIPS task ``REGRD.'' In performing the coordinate precession, we took
care to correct for the fact that the B1950 VLA phase calibrator
positions are given for equinox 1979.9 (e.g., 
Muxlow et al. 1995). 
The continuum image is included as an overlay in Figure~\ref{fig2}.

\section{Results}

Figure~\ref{fig1} shows the spectrum of the calibrated visibility data. The
spectrum is compatible with previous observations in the broadest
sense: there are many narrow, bright maser features spread from
 $v_{LSR} \sim -  $30 km s$^{ - 1} $ to  $\sim  $ 0 km s$^{ -
1} $. The spectrum does not agree in detail, however, owing to the
rapid variability of the masers associated with Ceph A; factor-of-two
variations occur on timescales as short as a few days 
(Rowland {\&} Cohen 1986). 

The distribution of maser spots relative to the position of the
continuum source HW2 is plotted in Figure~\ref{fig2}, and the
distribution of maser spots in both position and (grayscale-coded)
velocity is displayed in Figure~\ref{fig3}. Excluding the R5 maser
group, the spots spread over $\sim $ 1\arcsec\ east-west and $\sim $
0.5\arcsec\ north-south. Looking at the data qualitatively, there is a
broad trend of increasing velocity (blue-shifted to red-shifted) from
east to west. These results compare favorably with the previous VLA
observations of \water\ masers in HW2 (Torrelles, et al. 1996) and the
arcsecond-scale velocity gradient observed in SiO (2 $ \to $ 1)
(G\'{o}mez et al. 1999). The improved spatial and velocity resolution
of the MERLIN data, however, reveals finer structure than could be
mapped by the VLA. MERLIN resolves several arcuate groups of masers,
including the R2 -- R5 groups originally revealed by VLBA observations
(Torrelles, et al. 2001a; Torrelles, et al. 2001b).

Torrelles et al. (2001a) presented details of only the blue-shifted
masers associated with R4, and they measured a small proper motion
away from the center of curvature.  Accordingly, they interpreted the
U-shaped structure as a bow shock caused by an outflow originating
from a star near (in projection) to HW2 proper.  The conditions within
shocked molecular gas favor the generation of \water\ masers (Elitzur,
Hollenbach, {\&} McKee 1989; Kaufman {\&} Neufeld 1996), lending some
support to this scenario. The MERLIN observations reveal a structure
that is nearly ring-shaped (Figure~\ref{fig3}). The velocity gradient
of these brighter masers runs counter to the larger-scale gradient of
the HW2 region: velocities increase from the northwest to the
southeast along PA 140\degr\ (see the fitting analysis below and in
Table \ref{tab2}). This velocity gradient and elliptical geometry
suggest an alternative explanation: perhaps these R4 masers trace a
rotating ring of molecular gas surrounding the unseen protostar. The
ring is probably only an annular segment of a more extensive
circumstellar disk. In the following section, we evaluate these two
models in turn.

\section{Models for the R4 Masers}

\subsection{Model 1: Expansion into a Static Medium (Bow Shock)}
\label{subsec:shock}

The R4-A masers, those making up the northwestern arc of R4 (see
Figure~\ref{fig3}), present the most clearly defined structure of the
region. A bow shock would naturally explain the U-shape of the arcs
(e.g., Raga {\&} B\"{o}hm 1985; Raga 1986; Furuya et al. 2000) and the
expansion of the R4-A structure, but it is not clear that bow shocks
should produce the observed velocity gradient along the apparent major
(outflow?) axis.  The U-shape of a bow shock results from
edge-brightening, which for masers would require radial velocity
coherence for effective amplification.  The challenge for the bow
shock model, then, is to have the resulting line-of-sight velocity
coherence naturally provide both the arcuate shape of the maser spot
distribution and the observed radial velocity gradient along the
symmetry (outflow) axis (see Figure~\ref{fig3}).

We created a simple, parabolic shock front model to test this
scenario.  Following the formalism of 
Hartigan, Raymond, {\&} Hartmann (1987) and 
Hartigan, Raymond, {\&} Meaburn (1990), the geometry of the
shock front is given by:

\begin{equation}
\label{eq1}
z = \alpha \left( {x^2 + y^2} \right)
\end{equation}

\begin{figure*}
\plotone{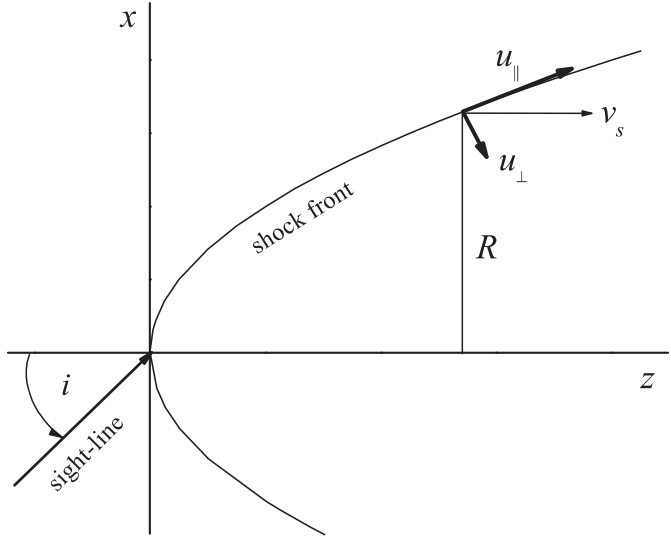}
\figcaption{A sketch of the coordinate conventions used to analyze the
bow shock model described in \S\ref{subsec:shock}.
\label{fig4}}
\end{figure*}

\noindent
where  $\alpha $ is a shape parameter,  $z $ is the position coordinate
along the outflow axis, and  $x $ and  $y $ are the position coordinates
orthogonal to the outflow axis; the sense of the coordinates is
illustrated in Figure~\ref{fig4}. As gas clouds fall into the shock, they are
accelerated only along the direction normal to the surface of the
shock. We will assume for this model that any bulk motions of the gas
clouds are negligible compared to their motion into the shock
front. To evaluate the kinematics of the maser spots, which trace the
post-shock gas, we need first to decompose the pre-shock velocity into
components perpendicular to the shock front and parallel to the shock
front. The post-shock velocities can then be estimated using the
Rankine-Hugoniot conditions. A more rigorous treatment would take into
account hydromagnetic effects. At this stage, however, the goal is to
evaluate coherence effects, and a more complex model is not
warranted. 




For the purpose of illustration, we assume isothermal (radiative)
conditions for the shock.  The actual conditions are probably closer
to isothermal than adiabatic, because, in order to produce \water\
masers, the pre-shock gas must already be dense, $n\left( \mbox{H}_2
\right) \ge 10^7 $ cm$^{ - 3} $ (Elitzur, et al. 1989; 
Kaufman {\&} Neufeld 1996).  
Based on the proper motion of
the maser arc (see the discussion in \S~\ref{subsec:model}), the shock
speed must be at least 13 km s$^{ - 1} $, and so the immediate
(adiabatic) post-shock temperatures are $\ga 10^{3} $~K 
(McKee {\&} Hollenbach 1980).  
Under these conditions, the cooling time is only a
few days, much shorter than the age of the maser arc. For the sake of completeness, it should be noted that an adiabatic shock model produces the same conclusions regarding radial velocity coherence; all that changes between the isothermal and adiabatic models is subtle, morphological details of the 
apparent structure of the shell. 



\noindent

In the reference frame of the observer (i.e., at rest with respect to the
stationary, pre-shock gas), the post-shock velocity is


\begin{eqnarray}
\label{eq4}
 {\rm {\bf {v}}} & = & {\rm {\bf {u}'}}_ \bot + {\rm {\bf {u}'}}_{{\rm {\bf 
\vert \vert }}} - v_S \,{\rm {\bf \hat {z}}} \nonumber\\ 
 & = & \frac{v_S }{1 + 4\alpha ^2R^2}\left[ {2\alpha \left( {x\,{\rm {\bf \hat 
{x}}} + y\,{\rm {\bf \hat {y}}}} \right) - {\rm {\bf \hat {z}}}} \right]
\end{eqnarray}

\noindent where {\bf u$'_{\bot}$} and {\bf u$'_{\vert\vert}$} are the
velocity components of the post-shock gas in a coordinate system
defined by the shock front (see Figure~\ref{fig4}), $v_{S}$ is the
shock speed, $\alpha$ is the shape parameter defined above, and $R =
\sqrt{x^2 + y^2}$. Thanks to the symmetry of the problem, we can take
the observer to be located at some distance away from the shock front
but confined to the $(x,z)$ plane (positioning in $y$ is equivalent to
an arbitrary rotation in PA). If the sight-line makes an angle $i$
with outflow axis such that $i = 0$ is a pole-on view, the radial
velocity of the post-shock gas is

\begin{equation}
\label{eq5}
{v}_R = v_0 + \frac{v_S }{1 + 4\alpha ^2R^2}\left( {2\alpha x\,\sin i - 
\cos i} \right)
\end{equation}

\noindent
where $v_{0} $ is the systemic velocity of the outflow source. For the
purposes of simulating the maser arc, we estimated the velocity and
shape parameters of the shock as follows. The maximum velocity of the
R4 maser spots is $\sim $ 7 km s$^{ - 1} $ relative to systemic. Again
assuming a fully radiative shock, deprojection gives the shock
velocity: $v_S \approx 7\sec i $ km s$^{ - 1} $ (in fact the proper
motion argues for a shock speed of about 13 \kms; however, the
magnitude of shock speed does not affect the conclusions). The
projected shape of the shock on the sky (${\alpha }') $ is related to
the intrinsic shape ($\alpha$) according to $\alpha = {\alpha }'\csc i
$ (Hartigan, et al. 1990). Based on a least-squares fit to the R4-A
maser spot positions, we find $\alpha \approx 44\,\csc i $(mas$^{ -
1}) $, which we used in our numerical models.

\begin{figure*}
\plotone{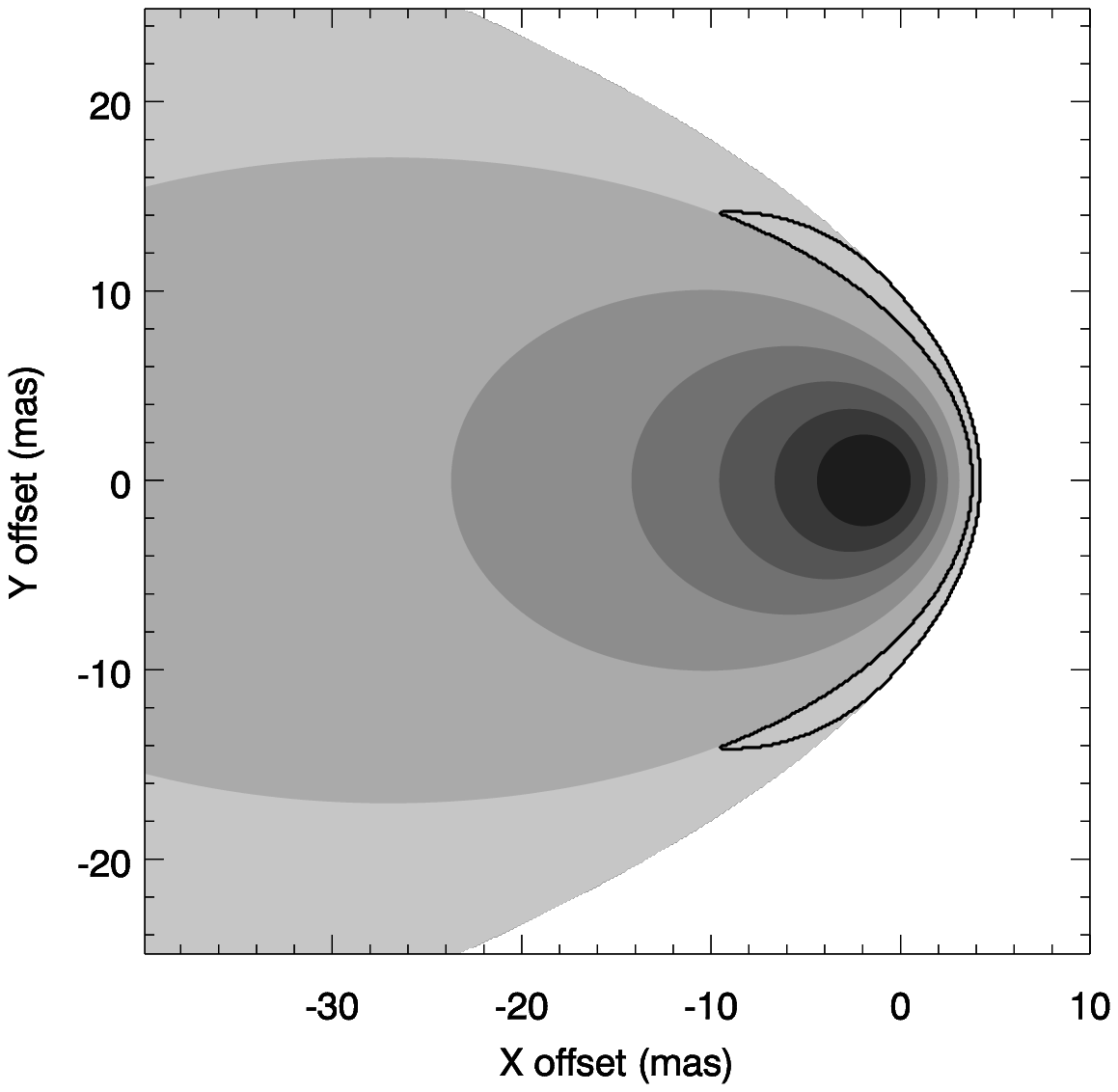}\\
\plotone{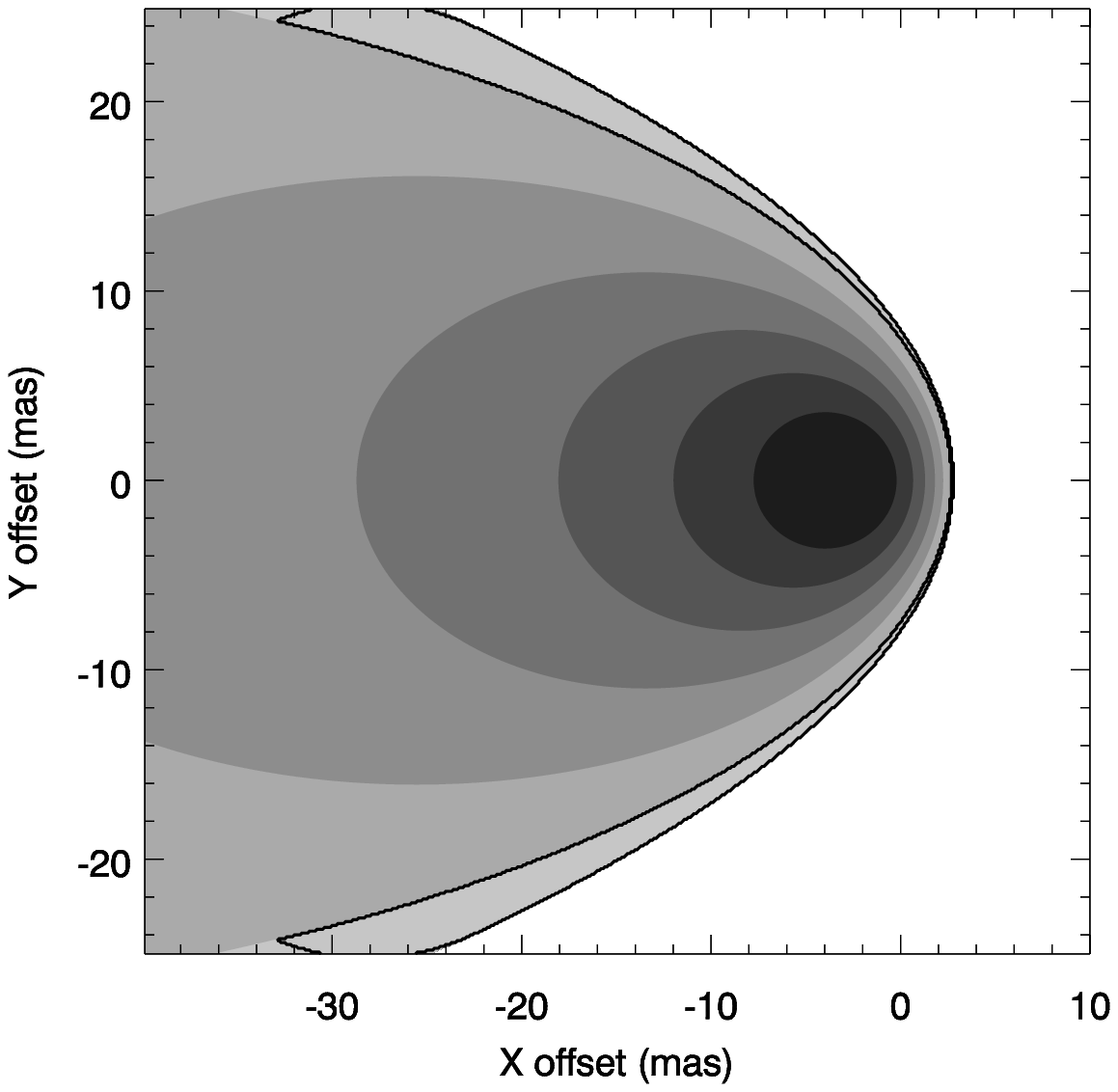}\\
\plotone{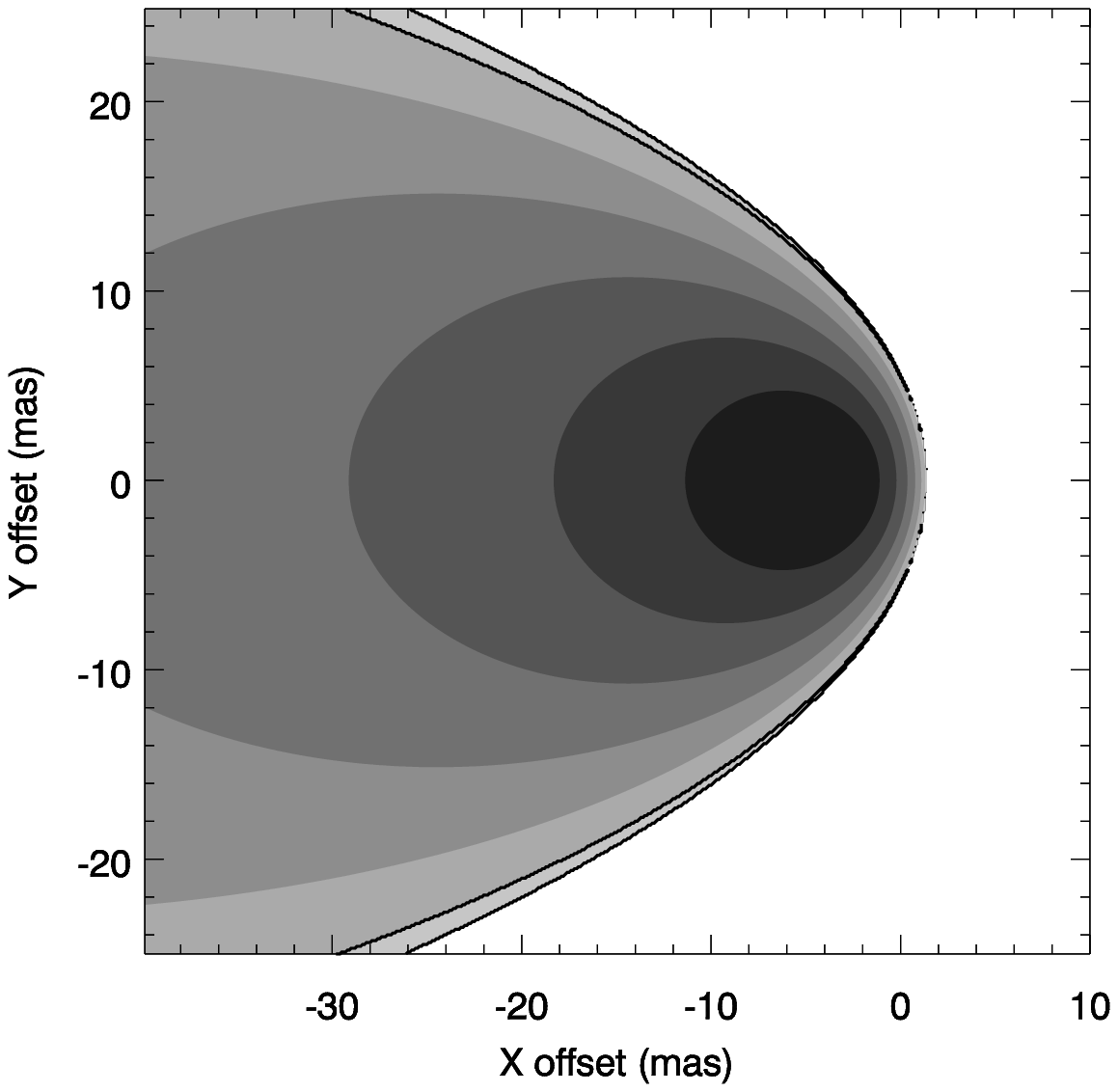}
\figcaption{Results of the bow shock model. From top to bottom, each panel
displays the bow shock model for inclinations  $i = $ 30\degr, 45\degr,
\& 60\degr, respectively. Grayscale contours represent the radial
velocity of the near side of the bow shock in steps of 1 km s$^{-1} $
from systemic (light gray) to $-8$ km s$^{-1} $ (black).  The heavy, dark
contour traces the region of the shock where the velocity difference
between the near-side and far-side is less than 2 km s$^{-1} $ and
where the Mach number exceeds unity.  Edge-brightening by velocity
coherence would produce a maser distribution similar in shape and
radial velocity as the region filling the heavy contours.
\label{fig5}}
\end{figure*}

We applied Equation~\ref{eq5} and the constraints on $v_S $ and
$\alpha $ to a set of models in an effort to mimic the properties of
the R4-A maser arc; the results are shown in Figure~\ref{fig5}. In
these models, the only free parameter is the inclination of the
outflow axis with respect to the line of sight. We assumed that masers
would arise only within the Mach angle (i.e., where the perpendicular
speed of the shock front is supersonic).  We further assumed that the
masers occupy a thin shell, appropriate for the short cooling
time. The masers are preferentially amplified when the velocity
difference between the near and far sides of the thin maser shell is
small.

The model maser regions plotted in Figure~\ref{fig5} are appropriate
for a maximum velocity difference of 2~km~s$^{-1} $, chosen to provide
a reasonable match to the spread of maser spots around the R4-A
arc. Relaxing either the thin shell assumption or the velocity
difference criterion broadens the projected shape of the maser region,
but the resulting maser spot velocity gradient along the outflow axis
would be unaffected. From inspection of Figure~\ref{fig5}, it is clear
that coherence effects naturally lead to edge-brightening and
therefore can explain the U-shape of the R4 maser distribution. The
$i=30\arcdeg $ model reasonably reproduces the extent of the R4-A
arc. On the other hand, this simple bow shock model cannot reproduce
the radial velocities of the maser spots. It seems that
edge-brightening preferentially amplifies regions with radial
velocities near the systemic velocity. In contrast, the maser spots
show a systematic gradient of $\sim 7 $~km~s$^{-1} $ along the
inferred outflow axis with the maximum velocities near the apex of the
U-shaped distribution. The kinematics of the maser spots are not
well-described by this simple model of a bow shock propagating into a
static medium, and it seems clear that the pre-shock gas must be
undergoing some systematic bulk motion to explain the velocity
gradient of the post-shock gas (i.e., the masers).

It is worth pointing out that changing the details of the shock model
do not significantly affect the results. In any model involving a bow
shock, the post-shock gas expands along and away from the outflow
axis. Sight-lines nearer the outflow axis intercept gas approaching
the observer on the near side and gas receding from the observer on
the far side. Velocity coherence is therefore poor near the outflow
axis. Sight-lines nearer the projected edge of the bow shock intercept
gas moving more nearly in the plane of the sky whether on the near or
far side. Radial velocities are therefore more coherent nearer the
projected edges of a bow shock and lie close to the systemic
velocity. 

\subsection{Model 2: Expansion into a Rotating Medium (Disk)}
\label{subsec:model}

This model expands on the previous one and assumes that the pre-shock
gas rotates around the source of an outflow or blastwave, presumably a
protostar or a newly formed star. Whether the outflow is collimated or
spherical, the rotational motion would be tangential to the strongest
part of the shock front. The post-shock gas therefore largely retains
the rotational component of motion and picks up an additional
expansion component. There remains, however, the issue of the arcuate
structure. Coherence effects only come into play if we observe the gas
very near the equatorial plane of rotation, but an arbitrary geometry
would not produce arcuate structure in this case. It seems likely,
then, that the masers arise from dense, post-shock clumps of gas,
without regard to larger-scale velocity coherence.

The simplest geometry that can produce the arcs is a spherical
shockwave expanding into an inclined, rotating disk. (Supposing an
asymmetric shockwave adds orientation parameters that we would be
unable to constrain.) The blastwave shock-heats molecular material in
the disk, leaving behind a temporary, patchy ring of maser emission
(Elitzur, et al. 1989). To estimate the properties of the maser
geometry subject to this blastwave and disk interpretation, we applied
a tilted-ring model akin to that used for HI galactic velocity fields
(Begeman 1989). To simplify the model and reduce the number of free
parameters, we made the following constraints or assumptions. (1) The
mass of the disk is much less than the mass of the central protostar,
and therefore the rotation curve is Keplerian. (2) The position angle
and inclination of the disk does not change significantly with radius
and time (that is, there is little warping or precession between the
1996 and 2000). (3) The ring expansion velocities may have changed
between epochs, perhaps as a result of mass-loading as disk material
is swept up. (4) The central protostar has not significantly
accelerated between epochs such that the systemic velocity of the disk
has not changed.  To accommodate these assumptions in practice, the
model comprises four inclined rings, one for each epoch (three for the
1996 VLBA observations).  Each ring is constrained to have the same
inclination and position angle, a common central mass to determine the
rotation speed of each ring, and two expansion velocities, one for the
1996 epochs and the other for the 2000 epoch. To be clear, the
expansion velocities are based on a purely kinematical fit to the
radial velocities of the masers. In the context of this model, the
expansion velocities measure post-shock velocities in the disk gas
rather than the motion of the blastwave proper.

Because the VLBA astrometry is referenced to a different sky position, we 
also separately fit the center positions of the VLBA and MERLIN data. 
Ultimately, we aligned the VLBA data to the MERLIN data assuming a common 
center for the R4 ring. This latter assumption, equivalent to an assumption 
of zero proper motion of the central protostar, does not affect the model 
per se, but it does provide an astrometric alignment between the data sets 
that can be checked for self-consistency. We will return to this astrometry 
check below.

\begin{figure*}
\plotone{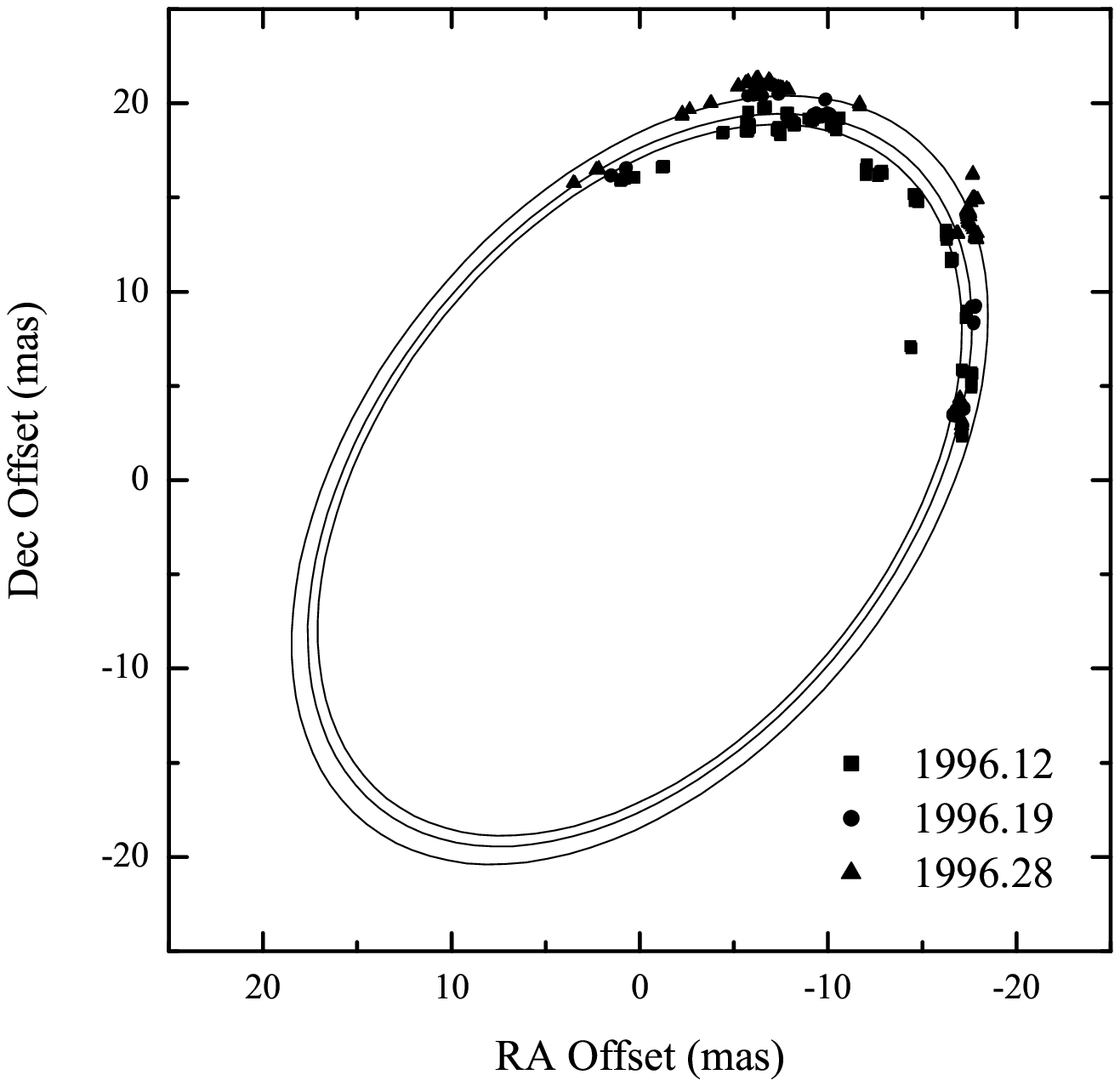}
\plotone{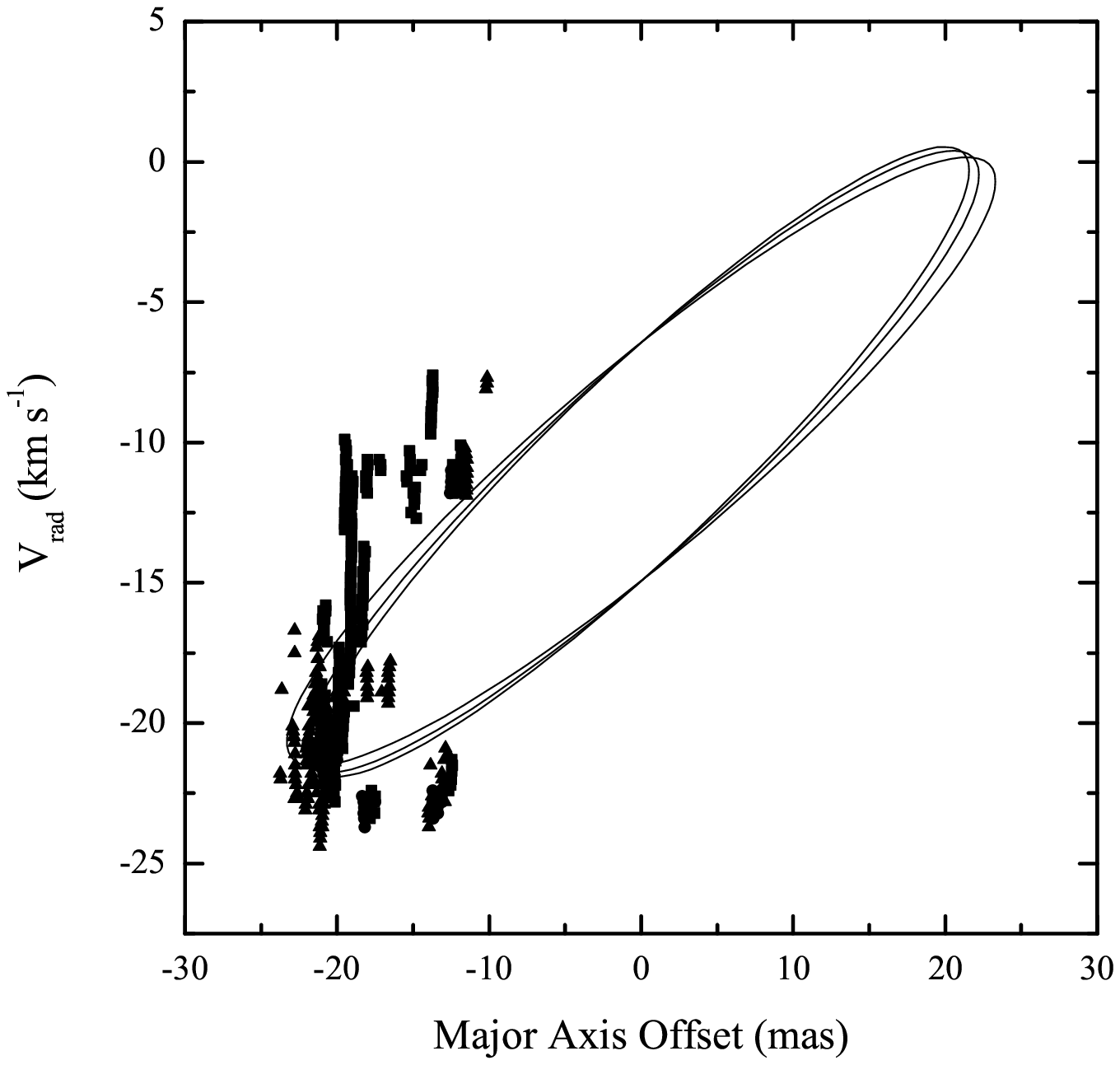}
\plotone{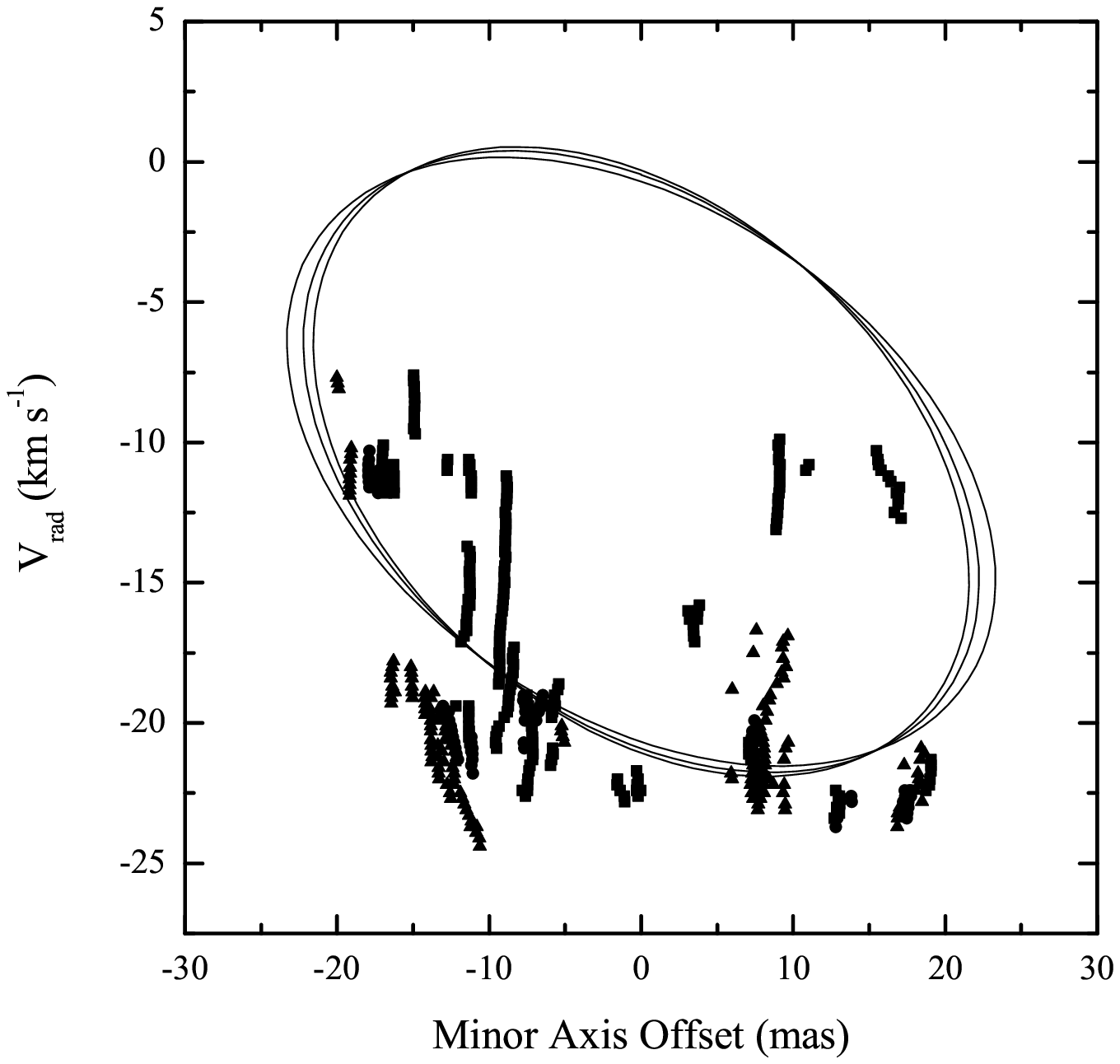}
\figcaption{Results of the rotating ring model for the 1996 VLBA observations of
(Torrelles, et al. 2001a). In each panel, the model is plotted as a
solid line. The VLBA data are plotted as symbols, and the type of
symbol depends on the epoch of observation as described in the legend
of the topmost panel.  Sky coordinates are referenced to the model
ring center (see Table 1).  {\em Upper Left Panel}: data and model projected
on to the plane of the sky. The best fit sky offset to the ring center
has been subtracted from the sky coordinates. {\em Upper Right Panel}:
Radial velocity vs. offset along the ring major axis. {\em Bottom
panel}: radial velocity vs. offset along the ring minor axis. Note
that the minor axis offset coordinates have been deprojected for ring
inclination.
\label{fig6}}
\end{figure*}

\begin{figure*}
\plotone{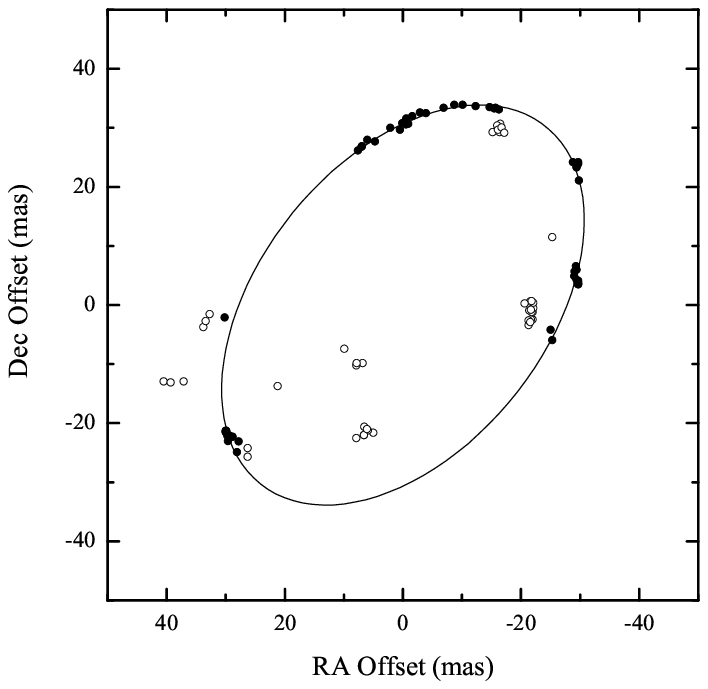}
\plotone{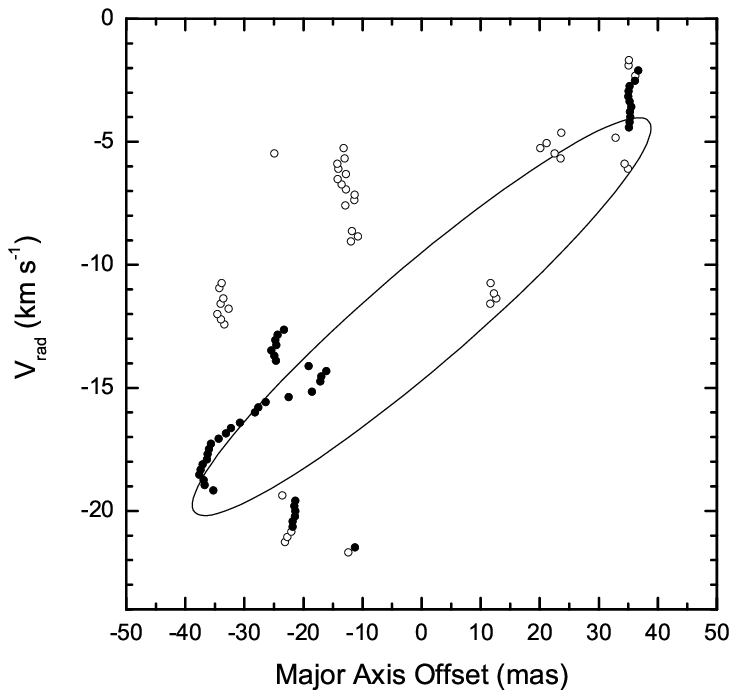}
\plotone{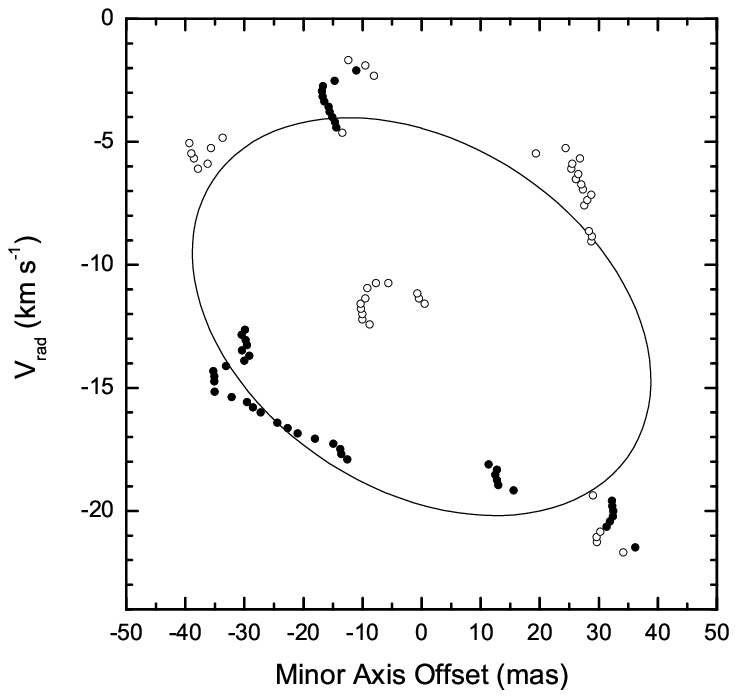}
\figcaption{Results of the rotating ring model for the 2000 MERLIN
observations. In each panel, the model is plotted as a solid
line. Open circles mark data that were rejected as not belonging to
the ring during the fitting process; solid circles mark data used in
the fit.  Sky coordinates are referenced to the model ring center (see
Table 1).  {\em Upper Left Panel:} data and model projected on to the plane
of the sky. The best fit sky offset to the ring center has been
subtracted from the sky coordinates. {\em Upper Right Panel:} Radial
velocity vs. offset along the ring major axis. {\em Bottom Panel:}
radial velocity vs. offset along the ring minor axis. Note that the
minor axis offset coordinates have been deprojected for ring
inclination.
\label{fig7}}
\end{figure*}

Fitting involved a non-linear least squares technique minimizing the
$\chi ^{2} $ difference between the model and data positions and
velocities.  Among the MERLIN data, many fainter maser spots clearly
do not associate with the arcuate structure traced so clearly by the
R4 masers. To remove these unassociated masers objectively, we
repeated the fitting procedure after clipping data for which the
deprojected radius disagreed with the best-fit ring radius by more
than 3 $\sigma $, where $\sigma $ was calculated by a quadratic sum of
the model radius uncertainty and the data positional uncertainty. The
results of this modeling for all free parameters are listed in Tables
\ref{tab1} and \ref{tab2}, and the sky and kinematical projections are
presented in Figures~\ref{fig6} and \ref{fig7} for the 1996 VLBA and
2000 MERLIN observations, respectively. The enclosed mass is 3.2 M
$_{\sun} $; a ZAMS star of this mass would be of spectral type A0 / B9
(de Jager {\&} Nieuwenhuijzen 1987; Palla {\&} Stahler 1993; Drilling
{\&} Landolt 2000).

\tabcaption{Results of the rotating ring model fit to the R4 maser
data. \label{tab1}}
\begin{center}
\begin{tabular}{lrcll}
\tableline
\tableline
Parameter& \multicolumn{3}{c}{Value} &  Units \\
\tableline
RA offset &  $ - 7.0 $&  $\pm  $& 0.1& mas \\
Dec offset  &  $ - 27.4 $&  $\pm  $& 0.2& mas \\
Inclination& 50&  $\pm  $& 1& degrees \\
PA& 142&  $\pm  $& 2& degrees \\
Sys. velocity  &  $ - 12.1 $&  $\pm  $& 0.5& km s$^{ - 1} $ \\
 $v_{out} $ (1996)& 5.5&  $\pm  $& 0.5& km s$^{ - 1} $ \\
 $v_{out } $(2000)& 5.3&  $\pm  $& 0.5& km s$^{ - 1} $ \\
Central mass& 3.2&  $\pm  $& 0.2& M $_{\sun} $ \\
\tableline
\end{tabular}
\end{center}
{{\em Table 1 Comments: }Position offsets refer to the location of the center of the ring
and are referenced to the position of the brightest maser: $\alpha
$(2000) = 22$^{\mathrm h} $ 56$^{\mathrm m} $ 17\fs9807, $\delta
$(2000) = +62\degr 01\arcmin 49\farcs429 (accurate to about 12
mas). The best fit radii are listed in Table~\ref{tab2}.}

\tabcaption{Best fit radii for the rotating ring model fits to the
R4 maser data. \label{tab2}}
\begin{center}
\begin{tabular}{lrclrcl}
\tableline \tableline Epoch& \multicolumn{3}{c}{Radius (mas)} &
\multicolumn{3}{c}{Radius (AU)} \\ 
\tableline 
1996.12& 21.6&  $\pm  $& 0.2& 15.7&  $\pm  $& 0.1 \\ 
1996.19& 22.2&  $\pm  $& 0.2& 16.1& $\pm  $& 0.1 \\ 
1996.28& 23.3&  $\pm  $& 0.2& 16.9&  $\pm  $& 0.1 \\
2000.27& 38.7&  $\pm  $& 0.1& 28.1&  $\pm  $& 0.1 \\ 
\tableline
\end{tabular}
\end{center}

Figures~\ref{fig6} and \ref{fig7} show that the expanding, rotating
ring models describe well the velocity gradient and spot distribution;
certainly, this disk interpretation provides a better match to the
maser spot kinematics than does the bow shock model.  The positions of
most of the maser spots agree to within a few mas of the model ring
positions. The average proper motion of the R4 arc, based on the increasing
radius of curvature between 1996 and 2000 (Table~\ref{tab2}), is $3.9\pm 0.2 $ mas yr$^{-1}
$ ( $13.3\pm 0.1 $~\kms). The radial velocities generally agree to
within 5 km s$^{ - 1} $. The scatter of the velocity residuals might
be attributed to substructure in the shockwave and turbulence in the
post-shock gas; such scatter is observed in masers associated with
outflows (e.g., Elitzur 1992).

There remain, however, significant outliers in both position and
velocity.  It is difficult to account for large positional outliers in
the context of this shockwave model. Masers lying within the ring
radius for a given epoch may arise in clumps with longer post-shock
relaxation times; masers outside the ring might be produced in a
radiatively heated shock precursor region (Tarter {\&} Welch 1986).
Either explanation is unfortunately speculative and difficult to test
owing to the small angular scale and high extinction of the region. On
the other hand, the outliers may arise from clumps not physically
associated with R4-A. Cepheus A is an active star forming region with
many radio sources (Sargent 1977; Hughes {\&} Wouterloot 1984b;
Hughes, Cohen, {\&} Garrington 1995); sight-lines to the R4 region may
well intercept more than one system of maser spots.

\begin{figure*}
\plotone{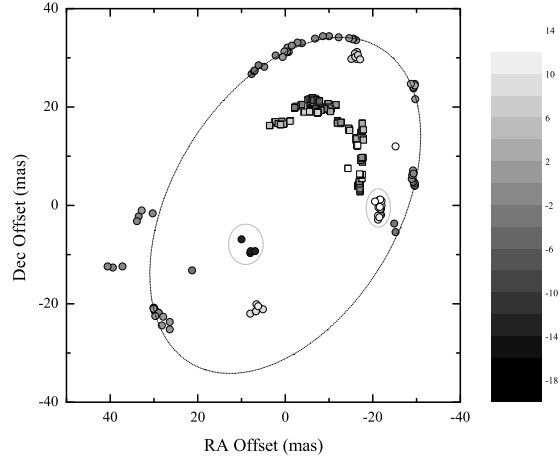}
\figcaption{Residual analysis of the rotating ring model. Sky coordinates are
referenced to the model ring center (see Table 1).  The symbols
marking the R4 maser positions are shaded according to their residual
velocity ( $v_{LSR}-\mbox{model} $ in \kms). Circles represent the MERLIN (2000)
data, and squares mark the VLBA (1996) positions of the R4-A masers.
The dotted line traces the best fit ring to the MERLIN data, and the
smaller, gray ellipses surround the two spot groups with the largest velocity
discrepancies with respect to the rotating ring model.
\label{fig8}}
\end{figure*}

We investigated whether the MERLIN positional outliers might yet arise
from the putative disk but at radii different from the shock ring. If
the outliers are part of the disk, then their radial velocities should
be compatible with the Keplerian rotation curve used to model the
rings. We calculated the theoretical radial velocities at the
positions of each of the R4 maser spots and subtracted the model
velocities from the observed velocities. The resulting residual field
is plotted in Figure~\ref{fig8}. The velocity residuals of maser spots
located on the rings are small, as would be expected from inspection
of Figures~\ref{fig6} and \ref{fig7}. More interestingly, the velocity
residuals of the maser spots not associated with rings are
substantially reduced, a result arguing that most of the maser spots,
whether on the ring for a given epoch or not, appear to be
kinematically associated with R4.

\begin{figure*}
\plotone{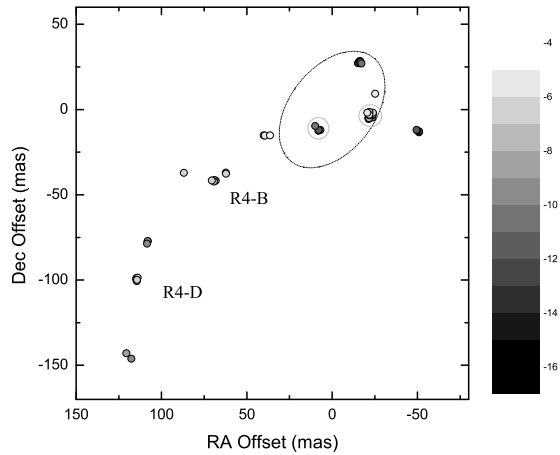}
\figcaption{The MERLIN positions of the R4 masers within the velocity range 15 km
s$^{-1} $  $< v_{LSR} < $ 5 km s$^{-1} $. Maser spots whose kinematics are
well described by the rotating ring model (velocity residuals  $< 5 $ km
s$^{-1} $) have been removed to emphasize the distribution of the
outliers. The spots are shaded based on their radial velocity. The
dotted line traces the best fit ring to the MERLIN data, and the
darker ellipses surround the two spot groups with the largest velocity
discrepancies with respect to the rotating ring model.
\label{fig9}}
\end{figure*}

Two groups of R4 masers are glaring exceptions; these groups have been
circled on Figure 8. Both groups have velocities that are too near
systemic for their projected position onto the model disk, and, as
such, show the largest velocity discrepancies with respect to the disk
model. Their velocities are in better agreement with the range of
velocities among the R4-B and R4-D masers, located southeast of the R4
rings and forming an apparent ``tail'' extending away from the R4 arc
(see Figure~\ref{fig3}). Figure~\ref{fig9} shows the kinematical
distribution of the outlier maser spots with radial velocities in the
range $ - 15\mbox{ km s}^{ - 1} \le \mbox{ }v_{LSR} \le -
5\,\,\mbox{km s}^{ - 1} $, i.e., the range of velocities spanned by
the R4-B and R4-D masers. It is interesting to see that, with the ring
masers removed, the outlier groups appear to align with a larger arc
traced by the R4-B and R4-D masers and extending through the R4
ring. We can only speculate whether or not this arc is somehow
associated with the R4 ring; it seems more likely that it is a
separate maser arc that happens to lie near the same sight-line to the
R4-A arc. Further observations may reveal whether these outlier
groups, if they persist, participate in a systematic proper motion
with the R4-B and R4-D masers.

\subsection{Discussion of the Disk Model}

The models described above are necessarily simplifications, but they
provide the basic measurements of the expansion of the R4 masers
needed to deduce some properties of the shockwave and the pre-shock
gas. The disk model is somewhat limited by the assumption of spherical
symmetry of the shockwave and cylindrical symmetry of the
disk. Lacking additional data for the source of the shockwave, one
might relax the condition of spherical symmetry and find a better fit
to the data. For example, a more collimated outflow inclined into the
plane of the disk might explain the lack of masers along the minor
axis of the R4 maser distribution. The R4 maser spots sample the local
kinematics unfortunately too sparsely to justify a more sophisticated
model. It seems nevertheless likely that the shockwave must at least
propagate directly along the projected major axis, and the properties
of the spherical model apply to the shockwave along that axis.


Referring to Table~\ref{tab2}, the length of the semi-major axis
expanded by 11 AU between 1996 and 2000, corresponding to an average
ring expansion speed of 13~\kms. The expansion speed over the two
months of VLBA observations is somewhat larger, roughly 30 -- 40~\kms\
based on the proper motion between the 1996.12 and 1996.28 epochs, and
so it seems the ring expansion has been decelerating. The cause of the
deceleration may simply be mass-loading, analogous to the snowplow
phase of a supernova shell, although scaled down in energy and
momentum.  

There are unfortunately insufficent data to fit even a uniform
deceleration model believably; the simplest models (uniform
deceleration and snowplow deceleration) have two free parameters, but
there are only four data points. Nevertheless, to get an estimate of
how much the expansion speed may differ from the 1996 -- 2000 average
speed, we fit a snowplow model (after Dyson \& Williams 1997),
appropriate for expansion into a dense medium, to the data in
Table~\ref{tab2}. We anchored the initial radius to 15.7~AU (the
best-ring radius at epoch 1996.12) and allowed the initial velocity to
vary. The best-fit initial velocity is $41.3\pm 0.7$~\kms, predicting
an epoch 2000.27 velocity of $7.2\pm 0.7$~\kms. These values should be
viewed with caution, of course, because the present data do not
sufficiently sample the ring expansion well enough to support the
snowplow model in particular. The snowplow model does however show
that the expansion speed at epoch 2000.27 may differ from the average
by nearly a factor of two.

Following the arguments of \S~\ref{subsec:shock}, the ring expansion
speed should be identically the shock speed. In the discussion to
follow, we scale the shock speed to the average ring expansion speed,
13~\kms, accepting that the instantaneous shock speed may be somewhat
lower in the later epochs. Otherwise, equating the ring expansion speed and the shock
speed is valid provided the cooling time does not significantly vary
with distance from the source of the shockwave. It has been proposed
that the conditions necessary for \water\ maser emission may be caused
either by fast, dissociative shocks (Elitzur, et al. 1989) or slow,
non-dissociative hydromagnetic shocks (Kaufman {\&} Neufeld 1996). We
can rule out dissociative shocks for R4, which require $v_S \ga
50\,\,\mbox{km s}^{ - 1} $.  Our concern now is whether slower shocks
through a protostellar disk could produce the observed high brightness
temperatures.

Kaufman {\&} Neufeld (1996) generated an array of shock models to
calculate the efficiency of slow, C-type hydromagnetic shocks. They
defined the efficiency as the ratio of the maser luminosity output
divided by the total mechanical energy provided by the shock:
$\varepsilon _{sat} = 2L_{sat} / A_m n_0 \mu v_S^3 $, where $L_{sat} $
is the luminosity of the (saturated) water maser, $A_m $ is the
surface area of the maser source, $n_0 $ is the pre-shock hydrogen
density, and $\mu = 4.2\times 10^{ - 24}\,\mbox{g} $ is the mean mass
per hydrogen atom. The maximum maser luminosity occurs where the
product $n_0 \varepsilon _{sat} $ is a maximum. Using $v_S =
13\,\,\mbox{km s}^{ - 1} $ and interpolating Figure~6 of Kaufman {\&}
Neufeld, the maximum luminosity occurs for $n_0 = 6.9\times
10^7\,\mbox{cm}^{ - 3} $, at which $\varepsilon _{sat} = 4.0\times
10^{ - 5} $. The predicted luminosity is then $L_{sat} = 3.9\times
10^{25}\,\left( {\ell / \mbox{AU}} \right)^2\,\mbox{erg s}^{ - 1} $
where $\ell $ is the size of the maser cloud. The masers are probably
beamed; we will make the simplifying assumption that the maser arises
from a cylinder of cross-sectional diameter $d $. Taking into account
the correction for cylindrical beaming, an observer in the path of the
beam would infer an isotropic luminosity:

\begin{equation}
\label{eq6}
L_{iso} = 3.9\times 10^{25}\left( {\frac{\ell }{AU}} \right)^2\left(
{\frac{\ell }{d}} \right)^2\,\mbox{erg s}^{ - 1}
\end{equation}

VLBA measurements limit $d < 0.4\,\mbox{AU} $ (Torrelles, et
al. 2001a), giving

\begin{equation}
\label{eq7}
L_{iso} > 2.4\times 10^{26}\left( {\frac{\ell }{AU}}
\right)^4\,\mbox{erg s}^{ - 1}
\end{equation}

The flux density of the brightest R4 maser spot is 650 Jy in a 0.2 km
s$^{ - 1} $ channel. Assuming isotropic emission, the luminosity of
any of the maser spots is therefore $L_{obs} \le 6.0\times
10^{27}\,\mbox{erg}\,\mbox{s}^{ - 1} $. The size of the maser cloud
required by the Kaufman {\&} Neufeld (1996) model is found by equating
the observed $\left( {L_{obs} } \right) $ and predicted $\left(
{L_{iso} } \right) $ maser luminosities: $\ell \le 2.2\,\mbox{AU}
$. It seems that the slow shocks model could plausibly produce the
observed maser luminosities insofar as the inferred path length is
less than 7{\%} the radius of the disk. For comparison, the gas scale
height derived assuming hydrostatic equilibrium (and neglecting
self-gravity) is $H\sim c_S R^{3 / 2} / \left( {GM} \right)^{1 /
2}\sim 5.4\,T_3 ^{1 / 2}\,\mbox{AU} $, where $T_3 = T / 1000\mbox{ K}$
(Frank, King, {\&} Raine 1992). Optical and infrared imaging of
circumstellar disks in a variety of environments measure scale heights
ranging from a few AU to tens of AU at comparable distances from the
central star (e.g., Beckwith {\&} Birk 1995; Stapelfeldt et al. 1998;
Heap et al. 2000; Stapelfeldt 2000; Stapelfeldt et al. 2000).

The parameters of the shock model, taking into account the properties
inferred from the Kaufman {\&} Neufeld (1996) model, are sufficient to
allow us to estimate the mass of the circumstellar disk.  Assuming
constant scale height, uniform density, and cylindrical symmetry, the
disk mass is:

\begin{equation}
\label{eq8}
M_{disk} \approx 0.97\left( {\frac{n_0 }{6.9\times 10^7\,\mbox{cm}^{ - 3}}} 
\right)\left( {\frac{R}{28\,\mbox{AU}}} \right)^2\left( 
{\frac{H}{2.2\,\mbox{AU}}} \right)M_ \oplus 
\end{equation}

\noindent where we have normalized $R $ to the radius of curvature
based on the MERLIN observations (Table~\ref{tab2}).  Note that the
pre-shock density is optimized for the most luminous maser emission,
as discussed above. Such a low disk mass is reminiscent of Vega-like
debris disks, whose masses are in the few $M_{\oplus}$ range (e.g.,
Sylvester \& Skinner 1996).  The disks surrounding young stars and
protostars are measured to be $\sim 0.01 {\mbox{-- a few}} M_{\odot}$
(e.g., Hillenbrand et al. 1992; Chandler \& Richer 1999 \& references
therein).  Whereas it is tempting to infer that the R4 system might be
more evolved than the surrounding protostars of the Ceph A region, it
seems just as likely that the gas associated with R4 is very
clumpy. Masers might then arise from relatively low density gas but
are quenched in higher density regions. Nevertheless, the mass
estimate is at least self-consistent with our original assumption of
negligible disk mass. As the ring expands and traces out the rotation
curve, future epochs of maser observations may place better dynamical
constraints on the disk mass.


Based on the kinematic model fitting, the expansion velocity of the R4
maser spots is roughly 5 km s$^{ - 1} $ (Table~\ref{tab1}), which we
interpret as the post-shock velocity. The difference between the shock
velocity and post-shock velocity is somewhat surprising: a strong
shock should produce post-shock velocities $3v_S / 4 \le {v}' \le v_S$
(McKee {\&} Hollenbach 1980; Dyson {\&} Williams 1997). Even allowing
for deceleration, the velocity difference remains an issue
for the 1996 epochs, during which the expansion speed is of order
40~\kms, eight times the post-shock speed. Weakening the shock to
$M\sim 1.5 $, referenced to the average shock velocity,
would drop the post-shock velocity to 5 km s$^{ - 1} $
but implies pre-shock temperatures $T > 10^4\mbox{ K}$ in order
to raise the sound speed sufficiently.  The pre-shock gas should be
molecular to produce post-shock masers (Elitzur, et al. 1989; Kaufman
{\&} Neufeld 1996), and so we can rule this option out.

Another explanation of the low post-shock velocity is that the
magnetic pressure of the post-shock gas may dominate the gas pressure
(see Liljestr\"{o}m {\&} Gwinn 2000 for an identical analysis of
W49N). Magnetic field pressure inhibits compression of the post-shock
gas, which, by conservation of momentum, increases the speed of the
gas leaving the shock. To the observer, then, the effect is to reduce
the post-shock velocity. In the limit where magnetic field pressure
dominates gas pressure, the magnetic pressure balances the ram
pressure of the pre-shock gas (Elitzur, et al.  1989; Kaufman {\&}
Neufeld 1996). The solution for the post-shock magnetic field strength
in terms of the pre-shock gas density $n_7 = n / 10^7\,\mbox{cm}^{ -
3} $ and shock velocity $v_S $ is:

\begin{equation}
\label{eq9}
{B}' = 31\left( {\frac{v_S }{\mbox{13 km s}^{ - 1}}} \right)n_7^{0.5} 
\,\mbox{mG}
\end{equation}

In the frame of the shock, the pre-shock magnetic field is given by
(Hollenbach {\&} McKee 1979; Liljestr\"{o}m {\&} Gwinn 2000):

\begin{eqnarray}
\label{eq10}
 B_\parallel & = & \frac{{u}'_ \bot }{v_S }{B}'_\parallel \nonumber\\ 
 & = & 19\,\left( {\frac{{B}'_\parallel }{{B}'}} \right)n_7^{0.5} \,\mbox{mG,} 
\\ 
\end{eqnarray}

\noindent
where parallel-line subscripts indicate components parallel to the
shock front. A magnetic field strength of $\sim 30 $~mG is plausible
in comparison with more direct measurements of magnetic field
strengths in similar astrophysical environments. For example, the
magnetic field strengths associated with the OH masers of Cepheus A
are a few mG (Wouterloot, Habing, {\&} Herman 1980; Cohen, Brebner,
{\&} Potter 1990).  Magnetic field strengths of 20 -- 100 mG have been
measured by observations of Zeeman splitting of \water\ masers found
in other star-forming regions (Fiebig {\&} Guesten 1989; Sarma,
Troland, {\&} Romney 2001).  Measuring Zeeman splitting of the H30
$\alpha $ recombination line, Thum {\&} Morris (1999) measured a field
strength of 22 mG in the corona of the circumstellar disk of MWC 349.

\section{Proper Motions of the R2, R3, and R5 Maser Regions}

The four-year separation between the VLBA and MERLIN observations
affords an opportunity to study proper motions over the entire HW2
region.  Unfortunately, the 1996 VLBA observations and the present,
2000 MERLIN observations are referenced to different maser spots, and
furthermore the 1996 VLBA data are not phase-referenced to a fixed
calibrator position. We can use, however, the proper motions of the R4
masers to calculate a bootstrapped astrometry between the data
sets. Based on the increasing radius of curvature of the R4-A masers
(and irrespective of the actual nature of the R4 masers, whether disk
or outflow or other), it seems clear that the R4-A region is expanding
away from some common center.

\begin{figure*}
\plotone{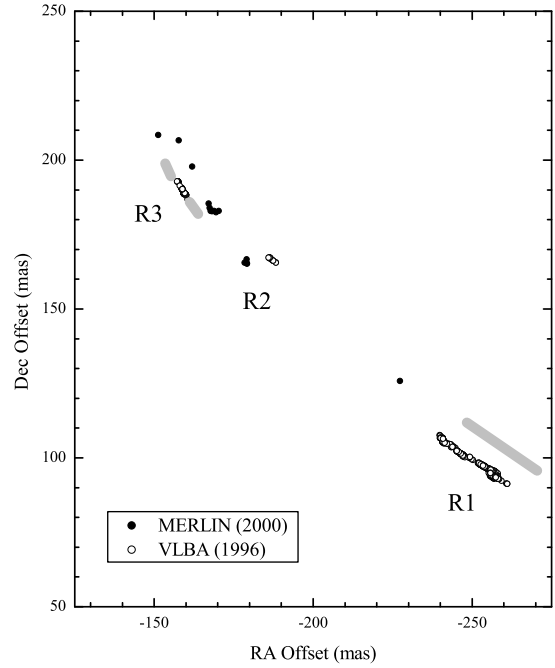}
\figcaption{Comparison of the MERLIN and VLBA data for maser regions R1, R2, and
R3. MERLIN detected no maser emission that could be clearly associated
with R1. The gray bands trace the predicted 2000 positions of the R1
and R3 masers based on the proper motion analysis of Torrelles et
al. (2001a).  The alignment between the MERLIN and VLBA data sets is
based on the assumption that the proper motion of the expansion center
of the R4-A masers is negligible.
\label{fig10}}
\end{figure*}

\begin{figure*}
\plotone{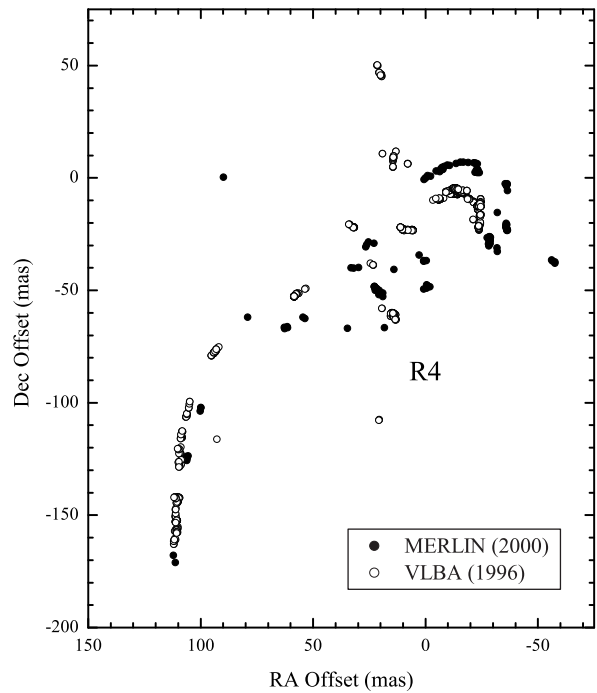}
\figcaption{Comparison of the MERLIN and VLBA data for the R4 maser regions. The
alignment between data sets is based on the assumption that the proper
motion of the expansion center of the R4-A masers is negligible.
\label{fig11}}
\end{figure*}

\begin{figure*}
\plotone{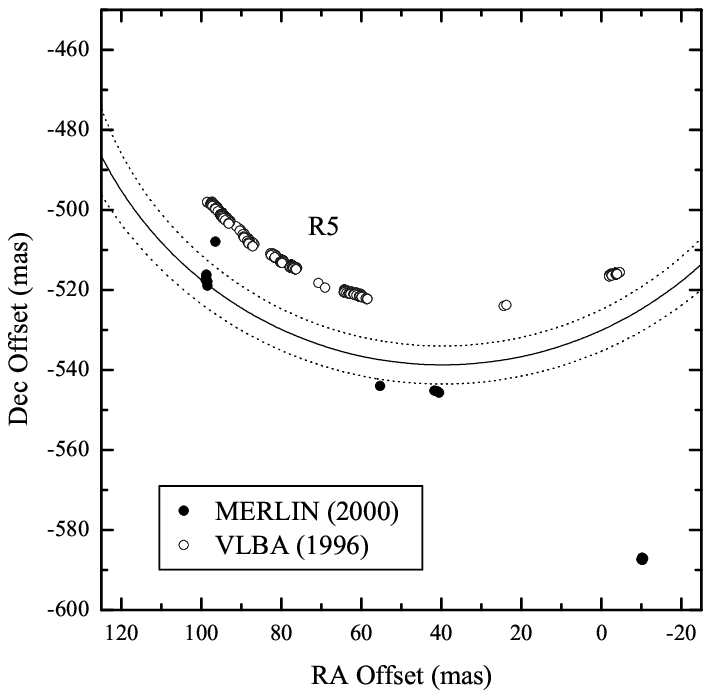}
\figcaption{Comparison of the MERLIN and VLBA data for the R5 maser region. The
alignment between data sets is based on the assumption that the proper
motion of the expansion center of the R4-A masers is negligible. The
VLBA data are plotted as open symbols, and the MERLIN data are plotted
as filled circles. The curved lines trace the region in which the
MERLIN data should appear based on the  $6.2\pm 1.8 $ mas yr$^{-1} $
proper motion measured from the three closely-spaced VLBA epochs. The
solid line traces the nominal, predicted location, and the dotted
lines mark the 90\% confidence prediction bands.
\label{fig12}}
\end{figure*}

As a first guess, we assume that proper motion of the R4 expansion
center is negligible. Figures~\ref{fig10}--\ref{fig12} plot the
relative positions of the MERLIN and VLBA data for the regions
R1--R5. The MERLIN positions of the R2 and R3 maser spots are
displaced by roughly 5 mas to the west of the positions predicted by
the proper motions measured by Torrelles et al.  (2001a), but the
declination alignment is somewhat better. From inspection of Figures 4
and 8 of Torrelles et al., the proper motion uncertainties are
probably of order 1 mas yr$^{ - 1} $, which propagates to $\sim $ 4
mas between the MERLIN and VLBA data sets. It is therefore unclear
whether the displacement might owe to proper motion of the R2 and R3
spots, proper motion of the R4 spots, which were used as the
astrometric reference, or both. Follow-up, phase-referenced MERLIN
observations should answer this question.

Torrelles et al.  (2001b) emphasized the circular symmetry of the R5
region and its apparent radial expansion over the 2 months
spanned by the VLBA observations. We re-measured the proper motion of the
R5 structure by fitting circles separately to each VLBA epoch. The R5
ring expands at a rate of 2.5  $\pm  $ 0.1 mas yr$^{ - 1} $ (8.6  $\pm  $
0.3 km s$^{ - 1}) $, and the expansion center moves at 1.4  $\pm  $ 0.1
mas yr$^{ - 1} $ (4.8  $\pm  $ 0.3 km s$^{ - 1}) $ towards P.A.  $126^
\circ \pm 6^ \circ  $.  For comparison, Torrelles et al. reported an
expansion speed of  $\sim  $9 km s$^{ - 1} $ and center motion of  $\sim  $
6 km s$^{ - 1} $ into PA  $\sim 143^ \circ  $. 

We calculated the location and geometry of the R5 arc as it should
have appeared for the (2000) MERLIN observations based on the
expansion measured by the (1996) VLBA observations; the results are
plotted in Figure~\ref{fig12}. Six out of the ten R5 maser spots detected by
MERLIN fall within the predicted range of sky offsets, and four are
displaced south of the prediction band by  $\sim 3 $ mas. We judge this
agreement to be very good, considering the uncertainty in predicting
four years of proper motion on the basis of two months of
monitoring. Based on the acceptable agreement for the R2, R3, and R5
masers, the original assumption for the astrometry bootstrap, namely,
negligible proper motion of the R4-A curvature center, seems
reasonable.

The argument can be reversed to constrain the proper motion of the R4
masers. Based on the  $\sim  $ 5 mas western displacement of the R2 and
R3 masers,  $\dot {\alpha } < 1.2\,\,\mbox{mas yr}^{ - 1} $. The
southern displacement of the R5 masers gives  $\dot {\delta } <
0.8\,\,\mbox{mas yr}^{ - 1} $. Taking these values as a limit on the
motion of the astrometric reference, the proper motion of the R4
expansion center must be less than  $\sim  $ 1.4 mas yr$^{ - 1} $ (5 km
s$^{ - 1}) $.

\section{Summary and Discussion}

The simplest explanation for the expansion of the R4 maser arcs
appears to be a ``slow,'' C-type hydromagnetic shockwave propagating
through a rotating, circumstellar disk surrounding a forming or young
A0 / B9 star.  Owing, in part, to the high extinction towards this
region, the central star has yet to be identified at any waveband (HW2
itself has only been identified in radio continuum). It is therefore
difficult to measure or constrain any other properties of the young
star and its associated shock except for what we have learned from the
maser kinematics.


The YSO IRAS~21391+5802 shows a similar, but smaller ($\sim 2$~AU
diameter) ring of masers, which presumably surround the young star
(Patel et al. 2000). In that source, the masers seem to occur at the
dust condensation radius of a radial outflow from the YSO. As the
molecular gas flows outward from the dust condensation front, the gas
cools, and the masers associated with that material fade out. The
maser ring persists as fresh material enters the dust condensation
front. This model predicts that the diameter of the maser spot
distribution should not change, however, in contrast to the appearance
of the R4 maser arcs, and furthermore the model predicts no strong
velocity gradient around the ring. The R4 maser arcs are furthermore
non-circular and much larger, extending well outside the dust
condensation radius (the masses of the two YSOs are similar). It seems
that, in contrast to IRAS~21391+5802, the R4 arcs cannot be explained
as a standing feature of a steady outflow; the maser properties are
better described by gas responding to a shockwave radially propagating
through a rotating medium.

The origin of the shockwave is unclear. One possibility might be
colliding spiral shocks, or other arcuate structure, forming as the
result of disk instabilities; for example, Durisen et al. (2001)
proposed this scenario to explain the occurrence of methanol masers in
protostellar disks. Disk and pattern rotation would alter the shape of
the maser arc over time, but it does not seem likely that spiral
shocks would produce the observed, outward propagation of the R4 maser
arcs, nor should they produce the increasing radius of curvature.
Spiral shock models also require that the disk is viewed nearly
edge-on for significant amplification (Maoz \& McKee 1998), but it
seems that the R4 arcs are viewed at an intermediate inclination.

A more promising explanation is provided by the work of Tscharnuter,
Boss, \& collaborators (Morfill, Tscharnuter, \& V\"olk 1985;
Tscharnuter 1987; Boss 1989). Using hydrodynamical simulations, they
have demonstrated that a collapsing cloud core can become unstable to
large oscillations owing to the thermodynamics of molecular hydrogen
dissociation and reassociation. As the protostar core forms, this
``hiccup'' instability can drive AU-scale, radial outflows of order
10~\kms\ (Boss 1989), comparable to the outward proper motion of the
R4 maser arcs. The hiccups recur as infalling material drives the
protostar again to instability. Balluch (1988) argued that this cycle
of instability proceeds for at least hundreds of years (i.e., at least
as long as the simulated duration of the numerical models). We
speculate that perhaps the disk associated with the R4 masers may be
responding to this sort of instability of the protostar. It is
difficult to evaluate the energetics involved based on the simulations
that have been published to date, and we are unaware of any work
modeling the impact of the hiccup instability on a surrounding
protostellar disk; such analysis is beyond the scope of the present
work. In addition, the hiccup models require fairly high infall rates,
which we do not have the data to justify in the case of the R4-A
masers. As was pointed out by Torrelles et al. (2001a; 2001b) in their
discussion of the neighboring maser source R5, the origin of disk
masers is yet poorly understood. Our shock interpretation argues for
more work towards understanding instabilities in protostars and the
impact of such instabilities on circumstellar disks.

\acknowledgements J. Gallimore and M. Thornley received travel support
from the National Radio Astronomy Observatory. R. Cool received
support from the National Science Foundation REU program, grant \#
0097424. We would also like to thank the staff of the Jodrell Bank
Observatory for their support. In particular, this work benefitted
from stimulating conversations with Anita Richards, Peter Thomasson,
and Tom Muxlow. Mark Claussen kindly sent us a pre-publication draft
of his 2001 review paper for IAU Symp. 206. An anonymous referee made
several very helpful suggestions that improved the text and pointed
out to us the properties of the maser ring of IRAS 21391+5802 (Patel
et al. 2000).

\end{document}